\documentstyle[preprint,tighten,aps]{revtex}
\begin{document}
\draft
\preprint{IFUP-TH 53/96}
\title{Topological charge on the lattice:\\
a field theoretical view of  the geometrical approach.
}
\author{Leonardo Rastelli, Paolo Rossi and Ettore Vicari}
\address{
Dipartimento di Fisica dell'Universit\`a and I.N.F.N.,
I-56126 Pisa, Italy
}
\maketitle
\begin{abstract}
We construct sequences of ``field theoretical'' 
lattice topological charge density operators which formally
approach geometrical definitions in
2-d $CP^{N-1}$ models and 4-d $SU(N)$ Yang Mills theories.
The analysis of these sequences of operators 
suggests a new way of looking at the geometrical method,
showing that geometrical charges can be
interpreted as limits of sequences of field theoretical
(analytical) operators.
In perturbation theory renormalization effects formally tend to vanish
along such sequences. But, since the perturbative expansion
is asymptotic, this does not necessarily lead
to well behaved geometrical limits. It indeed leaves open the 
possibility that non-perturbative renormalizations survive.

\end{abstract}



\section{Introduction}
\label{introduction}
The investigation of the topological properties of 4-d $SU(N)$ gauge
theories requires non-perturbative calculations. 
Lattice techniques represent our best source of non-perturbative
calculations. However investigating the topological properties of QCD 
on the lattice is a non-trivial task. 
Topology on the lattice is strictly trivial, and one 
relies on the fact that the  physical 
topological properties should be recovered in the continuum
limit. 

In the so-called field theoretical method~\cite{DiVecchia}, 
one constructs a local analytical function $q_L(x)$ of the lattice 
fields which has the topological charge density $q(x)$
as its classical continuum limit: 
\begin{equation}
q_L(x)\longrightarrow a^d q(x) + O(a^{d+1}).
\end{equation}
$q_L(x)$ is not unique, indeed infinitely many choices 
differing for higher order $O(a^{d+1})$ terms can be conceived. 
In this quite straightforward approach the major drawback is 
that related physical
quantities, such as the topological susceptibility, can only be obtained
after removing cut-off dependent lattice artifacts.
The classical continuum
limit must be in general corrected by including a renormalization
function~\cite{CDP},
\begin{equation}
q_{L}(x)\longrightarrow a^d Z(g_0^2) q(x) + O(a^{d+1}), 
\label{renorm}
\end{equation}
where $Z(g_0^2)$ is a finite function of the bare 
coupling $g_0^2$ going to one in the limit $g_0^2\rightarrow 0$.
The relation of the zero-momentum correlation of two $q_{L}(x)$ operators,
\begin{equation}
\chi_{L}= \sum_x \langle q_{L}(x)q_{L}(0) \rangle ,
\label{childef}
\end{equation}
with the topological susceptibility
$\chi_t$ is further complicated by an unphysical background term,
\begin{equation}
\chi_{L}(g_0^2)= a^d Z(g_0^2)^2\chi_t+M(g_0^2),
\label{chileq}
\end{equation}
which eventually becomes dominant in the continuum limit.
So in order to extract $\chi_t$ from this relation
one needs to evaluate $Z(g_0^2)$ and $M(g_0^2)$, which
is a hard task.

In order to overcome the problems caused by the above cut-off
effects, other methods have been proposed.
The so-called geometrical method~\cite{BergLuscher,Luscher} meets 
the demands that the topological charge on the lattice have 
the classical correct continuum limit and be an integer for every lattice
configuration in a finite volume with periodic boundary conditions.
In 4-d $SU(N)$ gauge theories this can be  achieved by 
performing an interpolation of the
lattice field, from which  the principal fibre bundle is reconstructed.
Since on the lattice each configuration can be continuously 
deformed into any other, integer valued geometrical definitions 
cannot have an analytical functional dependence on the lattice field. 
Due to their global topological 
stability, geometrical definitions should not be
affected by (perturbative) renormalizations.
On the other hand, as a drawback of their non-local nature
(leading to non-analiticity), they may be plagued by topological
defects on the scale of the lattice spacing, 
dislocations~\cite{Luscher2,Pugh}, whose non-physical contribution 
may either survive in the continuum limit (as 
in the $SU(2)$ gauge theory with Wilson action~\cite{Pugh}), 
or push the scaling region for the topological susceptibility
to large $\beta$ values.

Field theoretical and geometrical definitions 
may appear as two quite different approaches to the problem of
studying physical topological properties.
An interesting connection between them can be conceived by 
exploiting the following consideration.
Noting that geometrical definitions $Q_g$
can be written as a sum of local non-analytical terms,
$Q_g=\sum_x q_g(x)$, one can
construct sequences of field theoretical 
(analytical) definitions $q_L^{(k)}(x)$
formally approaching $q_g(x)$ as $k\rightarrow \infty$.
An example of such sequences  has been already given
for 2-d $CP^{N-1}$ models~\cite{CPN}.
 
The construction and analysis of these sequences 
lead to a new interpretation of the geometrical charges.
They can be considered as limits of appropriate sequences of
analytical operators, whose perturbative renormalizations 
formally tend to vanish along the sequences.
The asymptotic nature of the perturbative expansion,
which emerges from explicit calculations of the renormalizations,
does not necessarily lead to well behaved geometrical limits.
It indeed leaves open the possibility of a residual 
non-perturbative renormalization, which would manifest itself in
the contribution of lattice defects (the so-called dislocations) 
to the geometrical charge.
This interpretation may clarify the meaning of   
the geometrical charge from
a field theoretical point of view, and account for possible
discrepancies with other approaches.
Moreover, with increasing $k$, one may hope that
$q^{(k)}_L$ and their correlations enjoy a progressive suppression
of the renormalization effects, while 
for not too large $k$ their sensitivity
to lattice topological defects is reduced.

Our study is mainly done in the context of 2-d $CP^{N-1}$
models~\cite{Dadda-DiVecchia-Luscher,Witten}, which
are very useful theoretical
laboratories for testing non-perturbative numerical methods conceived
to study topological properties. 
We present an analysis based on large-$N$ and perturbative  
calculations, and numerical simulations. 
We then sketch an extension to QCD,
where we shall
construct a sequence of analytical topological charge density
operators approaching
L\"uscher's geometrical definition~\cite{Luscher}.

\section{Sequences of topological charge density operators
in  two-dimensional $\protect\bbox{CP^{N-1}}$ models}
\label{sec2}

\subsection{Topology in two-dimensional $\protect\bbox{CP^{N-1}}$ models}
\label{sec2sub1}

Two-dimensional $CP^{N-1}$ models are defined by the action
\begin{equation}
S= {N\over 2f} \int d^2x\,\overline{D_\mu z} D_\mu z
\end{equation}
where $z$ is a $N$-component complex scalar field subject to the 
constraint $\bar{z}z=1$,
and the covariant derivative $D_\mu =\partial_\mu +iA_\mu$ is defined 
in terms of the composite field
$A_\mu=i\bar{z}\partial_\mu z$.
Like QCD, they are asymptotically
free and present non-trivial topological structures
(instantons, anomalies, $\theta$ vacua). 
A topological charge density operator $q(x)$
can be defined,
\begin{equation}
q(x)={1\over2\pi}\varepsilon_{\mu\nu} \partial_\mu A_\nu,
\label{contch}
\end{equation}
with the related topological susceptibility
\begin{equation}
\chi_t = \int d^2x\langle q(x)q(0) \rangle.
\end{equation}
A pleasant feature of these models is the possibility
of performing a $1/N$ expansion. A large-$N$ analysis
of the topological susceptibility shows that $\chi_t=O(1/N)$.
One indeed finds~\cite{CPN-chi}
\begin{equation}
\chi_t\xi_G^2={1\over 2\pi N}\left( 1 -{0.3801\over N}\right)
+O\left({1\over N^3}\right),
\label{largeN}
\end{equation}
where $\xi_G$ is the second-moment correlation length.

Most lattice studies of 2-d $CP^{N-1}$ models
have been performed using the following actions:
\begin{equation}
S_L = -N\beta \sum_{n,\mu} |\bar{z}_{n+\mu}z_n |^2,
\label{standard}
\end{equation} 
where $\beta=1/(2f)$, $z_n$ is a $N$-component complex vector, 
constrained by
the condition $\bar z_nz_n = 1$, 
and
\begin{equation}
S_L^{(g)} = -N\beta\sum_{n,\mu}\left( 
   \bar z_{n+\mu}z_n\lambda_{n,\mu} +
   \bar z_nz_{n+\mu}\bar\lambda_{n,\mu} - 2\right),
\label{basic}
\end{equation}
where, beside the field $z$, a $U(1)$ gauge field $\lambda_{n,\mu}$ has 
been introduced, satisfying $\bar{\lambda}_{n,\mu}\lambda_{n,\mu}=1$.
The latter lattice formulation, $S^{(g)}_L$,
turns out to be particularly convenient for a large-$N$ 
expansion (see Ref.~\cite{CR_NC} for a review).

The original geometrical construction for the topological charge
proposed by Berg and  L\"uscher~\cite{Luscher} is 
\begin{eqnarray}
Q_g=&& \sum_n q_{n},\nonumber \\
q_{n}=&& 
{1\over 2 \pi} 
{\rm Im}\left[ \ln{\rm Tr}(P_{n+\mu+\nu}
P_{n+\mu}P_n)+ \ln{\rm Tr}
(P_{n+\nu}P_{n+\mu+\nu}P_n)\right],
\qquad \mu \neq \nu, 
\label{qgeomP}
\end{eqnarray}
where $P_{n} = \bar{z}_{n} \otimes z_{n}$ and the imaginary part of the 
logarithm is to be taken in $(-\pi, \pi)$.
For the lattice formulation (\ref{basic})
an alternative geometrical definition can be given 
in terms of the ``gauge'' field $\lambda_{n,\mu}$.
Introducing the plaquette operator
\begin{equation}
 u_{\lambda,n} =  \lambda_{n,\hat{1}} \, \lambda_{n+\hat{1},\hat{2}} 
\bar{\lambda}_{n+\hat{2},\hat{1}} \bar{\lambda}_{n,\hat{2}}, 
\label{uqt}
\end{equation}
one defines  $Q_{g,\lambda}$ by
\begin{eqnarray}
&&Q_{g,\lambda}=\sum_n q_{\lambda,n},\nonumber \\
&&u_{\lambda,n} = \exp(i 2 \pi  q_{\lambda,n}), 
\label{qgeomtheta}
\end{eqnarray}
where $q_{\lambda,n}\in (-\case1/2,\case1/2)$.
In view  of large-$N$ and perturbative calculations,
we write the infinite volume limit of $q_{\lambda,n}$
in the form
\begin{equation}
q_{\lambda,n} = {1\over4\pi}\varepsilon_{\mu\nu} (\theta_{n,\mu} + 
   \theta_{n+\mu,\nu} - \theta_{n+\nu,\mu} - \theta_{n,\nu}),
\label{qgeomtheta2}
\end{equation}
where $\theta_{n,\mu}$ is the phase of the field
$\lambda_{n,\mu}$, i.e.
$\lambda_{n,\mu} \equiv e^{i\theta_{n,\mu}}$.

On a finite volume and for periodic boundary conditions
$q_{n}$  and $q_{\lambda,n}$ generate integer values of the total
topological charge for each configuration.
They are not analytical functions of the lattice
fields $z$ and $\lambda$, and fail
to be defined on a zero measure set of ``exceptional'' 
configurations.
The main feature that makes these functions different from the ones
which we term ``analytical'' is the absence of single-valuedness.

Many Monte Carlo studies of the topological properties
of 2-d $CP^{N-1}$ models have been presented in the literature. 
A wide range of values of $N$ has been considered, both small
and large in order to test large-$N$ calculations. 
The present state of art is briefly summarized in the following. 

For $N=2$, that is for  the $O(3)$ non-linear
$\sigma$ model, recent simulations 
using the so-called classical 
perfect action~\cite{BBHN,DFP} and for a relatively large
range of values of $\beta$
seem to favour what suggested by semiclassical arguments,
that is  that $\chi_t$ would not be a physical quantity 
for this model,
in that a non-removable ultraviolet divergence
affects the instanton size distribution.
On the lattice this would manifest itself
in the fact that lattice estimators of $\chi_t$ 
do not properly scale approaching the continuum limit.
Earlier claims of scaling~\cite{O3} were probably originated
by the relatively small range 
of values of $\beta$ considered, and therefore scaling violations 
could easily be hidden behind the statistical errors.

For $N>2$ $\chi_t$ should be a physical quantity.
At $N=4$ the apparent failure of the geometrical method
is probably caused by a late approach to scaling of
the geometrical
definition (\ref{qgeomP}) when using standard lattice
actions such as (\ref{standard}) and (\ref{basic})~\cite{Wolff,HM,IM,CPN}.
Evidences of scaling and consistent results have 
instead been obtained by employing the field theoretical
and cooling methods using action (\ref{basic}) and its
Symanzik improvement~\cite{CPN},  and a  
geometrical method in the context of the classical perfect
action approach~\cite{Burkhalter}.

At larger-$N$, $N\gtrsim 10$, the geometrical estimator
(\ref{qgeomP}) shows scaling already at reasonable
values of the correlation length using both
action (\ref{basic}) and its Symanzik improvement~\cite{CPN,CPN2}.
The available Monte Carlo data at $N\gtrsim 10$
show that the dimensionless quantity $\chi_t\xi_G^2$ approaches,
although slowly,
the large-$N$ asymptotic behavior (\ref{largeN}), and 
quantitative agreement (within a few per cent of statistical errors)
is found at $N=21$.

In the following we shall construct sequences of operators
approaching the geometrical definitions (\ref{qgeomP}) and
(\ref{qgeomtheta}), and study their main features.

\subsection{Sequences of operators approaching geometrical definitions}
\label{sec2sub2}

To begin with, we consider a sequence of analytical operators
$q_{\lambda,n}^{(k)}$ approaching the geometrical
definition $q_{\lambda,n}$ expressed in terms of $\lambda_{n,\mu}$
fields, and within lattice formulations containing
explicitly the gauge field, like (\ref{basic}).
By taking appropriate combinations of the plaquette operator
$u_{\lambda,n}$ (cfr. Eq.~(\ref{uqt})), one can define a
sequence of local operators $q_{\lambda,n}^{(k)}$ 
which differ from $q_{\lambda,n}$ by higher and
higher order terms~\cite{CPN}
\begin{eqnarray}
q_{\lambda,n}^{(k)}  &&={1\over 2\pi}
  \sum_{l=1}^{k}\, \frac{(-1)^{l+1}}{l}\,
{ 2k \choose k-l }\, \frac{2}{{2k \choose k }}\, {\rm Im}(u_{\lambda,n})^l
\nonumber \\
&&= q_{\lambda,n} - \frac{k!^2}{(2k +1)!}
(2 \pi)^{2k} q_{\lambda,n}^{2k+1} +O(q_{\lambda,n}^{2k+3}).
\label{qkresto}
\end{eqnarray}

Similarly, one may define
\begin{eqnarray}
u_{1,n} & =  & {{\rm Tr}(P_{n+\hat{1}+\hat{2}}P_{n+\hat{1}}P_{n})\over
| {\rm Tr}(P_{n+\hat{1}+\hat{2}}P_{n+\hat{1}}P_{n})|},\nonumber \\
u_{2,n} & =  & {{\rm Tr}(P_{n+\hat{2}}P_{n+\hat{1}+\hat{2}}P_{n})
\over | {\rm Tr}(P_{n+\hat{2}}P_{n+\hat{1}+\hat{2}}P_{n})|},
\end{eqnarray}
and the quantities $q_{j,n}$ ($j=1,2$) by
\begin{equation}
u_{j,n} = \exp (i 2 \pi  q_{j,n})
\end{equation}
($q_{j,n}\in (-\case1/2,\case1/2)$), so that one can write
\begin{equation}
q_{n}= q_{1,n}+q_{2,n}.
\end{equation}
A sequence of operators approaching the geometrical definition 
(\ref{qgeomP}) can then easily constructed by
\begin{eqnarray}
q_n^{(k)} &&= q_{1,n}^{(k)}+q_{2,n}^{(k)},\nonumber \\
q_{j,n}^{(k)}  &&= {1\over 2\pi}\sum_{l=1}^{k}\, \frac{(-1)^{l+1}}{l}\,
{ 2k \choose k-l }\, \frac{2}{{2k \choose k }}\, {\rm Im}(u_{j,n})^l
\nonumber \\
&&= q_{j,n} - \frac{k!^2}{(2k +1)!}
(2 \pi)^{2k} q_{j,n}^{2k+1} +O(q_{j,n}^{2k+3}).
\label{qkresto2}
\end{eqnarray}
Strictly speaking these functions are not analytical everywhere
(they are not polynomials in the fields), but preserve the property of
single-valuedness.

Once the above definitions are given,
one can consider sequences of lattice topological susceptibilities,
either in terms of $\lambda$ or  $z$ fields:
\begin{equation}
\chi^{(k)} \equiv \sum_{n} \langle q_n^{(k)} q_0^{(k)}\rangle = 
\lim_{p^2 \rightarrow 0}
\langle\widetilde{q}^{(k)}(p)\widetilde{q}^{(k)}(-p) 
\rangle. \label{suscrossik}
\end{equation}
Under some general assumptions,
which essentially amount to assume the applicability
of an OPE to the correlation 
$q_L(x)q_L(y)$ when ${x\rightarrow y}$,
the relation between 
$\chi^{(k)}$ and the continuum topological susceptibility 
$\chi_t$ is
\begin{equation}
\chi^{(k)}(\beta) = a^2 Z^{(k)}(\beta)^2\chi_t + M^{(k)}(\beta),
\label{mixingsusccpn}
\end{equation}
where $M^{(k)}$ can be written in terms of the
identity $I$ and the trace of the energy-momentum tensor 
operator $T\equiv\frac{\beta(f)}{2f^2}\overline{D_\mu z}D_\mu z$, 
plus higher-order terms in $a$~\cite{CDPV},
\begin{equation}
M^{(k)}(\beta)= P^{(k)}(\beta)\left<I\right>  + a^2  
A^{(k)}(\beta) \left<T \right> \ + \ O(a^4).  
\label{mixingsusccpn2} 
\end{equation}
The expression of the background term $M$ in terms
of the identity and of the trace of the energy-momentum tensor 
(which are the only RG invariant operators of lower or
equal dimensions sharing the same quantum numbers of $\chi_t$) 
may be obtained by comparing the OPE's
of $q_L(x)q_L(y)$  and  $q(x)q(y)$,
and by taking their difference.

\subsection{Large-$\protect\bbox{N}$ and perturbative analysis}
\label{sec2sub3}

The sequence $q_{\lambda,n}^{(k)}$ is particularly suitable for
analytical calculations, which will show the main features of 
such constructions.
The reasons that make things easier are the following.
In both $1/N$ and perturbative expansion the variable
$\theta_{n,\mu}$ (defined by $\lambda_{n,\mu} 
\equiv e^{i\theta_{n,\mu}}$)
is used as a fundamental field, whose propagator 
can be  easily derived~\cite{CR_NC}. 
In the infinite volume limit the expression of the
geometrical charge density $q_{\lambda,n}$
is linear in terms of $\theta_{n,\mu}$ 
(cfr. Eq.~(\ref{qgeomtheta2})), therefore
one can easily obtain also the propagator of $q_{\lambda,n}$.
The use of the $q_{\lambda,n}$ propagator  
simplifies considerably the
study  of the large-$N$ behavior of the renormalization
effects in the sequence $q_{\lambda,n}^{(k)}$,
which can be written 
as polynomials in $q_{\lambda,n}$ (cfr. Eq.(\ref{qkresto})).
Indeed the leading non-trivial orders
of the corresponding  
$Z_\lambda^{(k)}(\beta)$, $P_\lambda^{(k)}(\beta)$, 
and $A_\lambda^{(k)}(\beta)$ 
can be obtained by evaluating tadpole-like diagrams,
whose lines are the $q_{\lambda,n}$ propagators,
and whose
vertices are the coefficients of the powers of $q_{\lambda,n}$
in Eq.~(\ref{qkresto}).
This occurs also for the evaluation of the leading
non-trivial order in standard perturbation theory.

In the large-$N$ limit one can unambiguosly identify the terms 
associated to $Z^{(k)}(\beta)$, $P^{(k)}(\beta)$, and $A^{(k)}(\beta)$.
It has been explicitly demonstrated that
$Q_{g,\lambda}$ and the corresponding susceptibility 
$\chi_\lambda$ in the infinite volume limit 
are not subject to renormalizations 
in the large-$N$ limit~\cite{DMNPR}, i.e. 
\begin{equation}
\chi_{\lambda} = \sum_n q_{\lambda,n} q_{\lambda,0} 
\rightarrow a^2 \chi_t
\label{chil}
\end{equation}
in the continuum limit (the same has been shown for the
definition (\ref{qgeomP}) too).
Using Eqs.~(\ref{qgeomtheta}) and (\ref{chil}),
one can deduce the following relationship valid in the lowest
non-trivial order of $1/N$ expansion
\begin{equation}
\chi^{(k)}_\lambda\equiv \sum_{n} 
\langle q_{\lambda,n}^{(k)} q_{\lambda,0}^{(k)}\rangle  
\approx a^2 \chi_t 
\left[1+2(2k+1)\alpha_k \big<q_{\lambda,0}^{2k}\big>\,
\right]+\alpha_k^2\, \sum_n
\big<q_{\lambda,n}^{2k+1}q_{\lambda,0}^{2k+1}\big>  \label{pippo}
\end{equation}
where  $\alpha_k = - (2 \pi)^{2k} \frac{k!^2}{(2k+1)!}$.

An asymptotic expansion in powers of the large-$N$ mass~\cite{CR_NC}, 
$m_0$, allows one
to separate the mixing with the identity from the 
term of dimension two (i.e. proportional to $a^2$)
in Eq.~(\ref{mixingsusccpn}), which contains the physical signal.
This is achieved by expanding the
$q_{\lambda,n}$ propagator in powers of $m_0$:
\begin{equation}
\langle \widetilde{q}_{\lambda}(p) \widetilde{q}_{\lambda}(-p)
\rangle = D_0 + m_0^2 D_1 + O(m_0^4).
\label{expans}
\end{equation}
Using results of Ref.~\cite{CR_NC}, the calculation 
of the functions $D_0$ and $D_1$ is quite  straightforward.
Then by comparing Eq.~(\ref{pippo}) with Eqs.~(\ref{mixingsusccpn})
and (\ref{mixingsusccpn2}), one can write down explicit analytical
expressions for $P_\lambda^{(k)}(\beta)$, $Z_\lambda^{(k)}(\beta)$, 
and $A_\lambda^{(k)}(\beta)$
in the lowest non-trivial order of $1/N$~\cite{Rastelli}. 
In the following we shall discuss some of their properties.
We shall not report their complete expressions because 
they would not be very illuminating.

The zero dimensional mixing with the identity, which dominates
the Monte Carlo signal in the continuum limit, turns out to be
\begin{equation}
P_\lambda^{(k)}(\beta) = O \left( \frac{1}{N^{2k+1}} \right).
\end{equation}
The renormalization functions corresponding to the terms
of dimension two turns out to be
\begin{equation}
Z_\lambda^{(k)}(\beta) = 1 + O \left( \frac{1}{N^k} \right),
\end{equation}
and
\begin{equation}
A_\lambda^{(k)}(\beta)= O \left( \frac{1}{N^{2k+1}} \right).
\end{equation}
In order to derive the lowest non-trivial order contribution
to $A_\lambda^{(k)}(\beta)$, we have used the fact
that in the $1/N$ expansion $\left< T \right>=O(1)$, 
indeed one finds~\cite{CR_NC}
\begin{equation}
\left< T \right>\xi_G^2={1\over 3\pi}+O\left( \frac{1}{N}\right).
\label{largeNt}
\end{equation}

In the $\beta \rightarrow \infty$ limit the large-$N$
expressions of $P_\lambda^{(k)}(\beta)$, $Z_\lambda^{(k)}(\beta)$, 
and $A_\lambda^{(k)}(\beta)$ 
reduce to formulas that could have been obtained by
standard weak-coupling perturbation theory. The correspondence
between large-$N$ expansion and standard perturbation theory is
obvious for $Z^{(k)}(\beta)$ and $P^{(k)}(\beta)$: it suffices 
to recognize that for
$m_0 = 0$ and $\beta \rightarrow \infty$ the large $N$ $\theta$-propagator
reduces to the corresponding perturbative propagator.
This correspondence has been explicitly verified also for
$A^{(k)}(\beta)$. One obtains
\begin{equation}
Z^{(k)}(\beta) = 1 -k!\,\left(\frac{1}{\beta N}\right)^k +
 O\left[\left(\frac{1}{\beta N}\right)^{k+1}\right],
\label{Zkpertrossi}
\end{equation}
and for large $k$
\begin{eqnarray}
P^{(k)}(\beta) & \simeq & \frac{k!^4 (4k+2)!}{(2k+1)!^3} \,
\left( \frac{1}{\beta N} \right)^{2k+1} + 
O\left[\left(\frac{1}{\beta N}\right)^{2k+3}\right]  
\sim (2k)! \, \left( \frac{1}{\beta N}\right)^{2k+1},
\label{Pkpertrossi}\\
A^{(k)}(\beta) & \sim &  (2k)! \, \left( \frac{1}{\beta N}\right)^{2k+1}. 
\label{Akpertrossi}
\end{eqnarray}

Equations (\ref{Zkpertrossi}-\ref{Akpertrossi})
show the mechanism of systematic improvement of the local
operators $q^{(k)}_\lambda$. 
As $ k \rightarrow \infty $, $q_\lambda^{(k)} 
\rightarrow q_{\lambda}$
and the lowest-order renormalizations are proportional to
higher and higher powers of
$1/\beta$. However, the coefficients of the leading non-trivial
term grow so fast with $k$ that the convergence to
zero of $Z^{(k)}(\beta)-1$, $P^{(k)}(\beta)$ and $A^{(k)}(\beta)$  
cannot be uniform for $\beta \rightarrow \infty$. 
This fact leaves open the possibility 
that for fixed  $\beta$ some non-perturbative
 renormalization effects may
eventually survive as $k \rightarrow \infty$, i.e., as the
sequence approaches the geometrical definition.

Inspired by this phenomenon,
we suggest the following general picture.
The geometrical charge can be interpreted as the limit of a sequence of
field theoretical (analytical) operators. As far as (at a given $\beta$)
the renormalization effects tend to vanish along the sequence, the
geometrical object provides a well-behaved 
lattice estimator of the topological charge. If on the contrary 
the renormalization effects do not disappear for $k \rightarrow \infty$,
some pathology should arise in the geometrical method, such as 
the contribution of short-distance topological defects.
The asymptotic nature of the perturbative
expansion, which manifests itself in the growing of the
corresponding coefficients, leaves open the possibility
of a background term behaving, for example, as
$\sim \exp \left( - c\beta\right)$, which does not get
suppressed in the limit $k\rightarrow \infty$.
According to the value of $c$, either this term is suppressed in the
continuum limit, or it survives and spoils the expected asymptotic 
behavior  of $\chi_t$,  which should behave as a 
quantity of dimension $d$.
The absence of non-perturbative effects at finite $\beta$ 
would be guaranteed by  a convergent perturbative expansion
of the renormalizations.
A good continuum limit of the geometrical definition
would be assured by a perturbative expansion in which
the limit $k\rightarrow\infty$ commutes
with the continuum limit $\beta\rightarrow\infty$.

The calculation of the renormalizations
is more involved for the sequence expressed in terms
of the $z$ fields, but the conclusions are qualitatively the same.
In particular, since the expansion of $q_n$ in terms of 
perturbative fields (for example in Valent's parametrization~\cite{Valent})
 starts with a bilinear term, 
it is easy to see  from Eq.~(\ref{qkresto}) that $Z^{(k)}(\beta) -1$ and 
$P^{(k)}(\beta)$ will get the leading
contribution from diagrams respectively with $2k$ and $2k+2$ loops.
Therefore
\begin{equation}
Z^{(k)}(\beta) = 1 + O\left[\left(\frac{1}{N\beta}\right)^{2k} \right],
\label{zeta}
\end{equation}
and
\begin{equation}
P^{(k)}(\beta) = O\left[ \left(\frac{1}{N\beta}\right)^{2k+2}\right].
\label{pp}
\end{equation}
These results hold for both lattice actions (\ref{standard}) 
and (\ref{basic}).

\subsection{Numerical analysis by heating method}
\label{sec2sub4}

Since estimators involving the field $\lambda_{n,\mu}$
are subject to large fluctuations in Monte Carlo  simulations,
in our numerical analysis we shall consider the sequence
approaching the geometrical definition in terms of $z$ fields
only, i.e. $q_n^{(k)}$ defined in Eq.~(\ref{qkresto2}).
We present a numerical 
investigation of the corresponding renormalization
effects  by
using the so-called heating method~\cite{Teper,heating}
(see Ref.~\cite{CPN} for an implementation of this method to
2-d $CP^{N-1}$ models).
The study of the renormalization effects by heating
method relies on the possibility of somehow separating 
the various contributions in Eq.~(\ref{mixingsusccpn})
in off-equilibrium simulations.
This is made possible when they are originated  by lattice modes
which behave differently under local thermalization
(large vs small scale modes, Gaussian vs
topological modes). 
Assuming such a distinction of lattice modes,
estimates of the multiplicative renormalization 
$Z^{(k)}$ are obtained by measuring
$Q^{(k)}=\sum_n q_n^{(k)}$ on ensembles 
of configurations constructed by heating an instanton-like
configuration (carrying a definite topological charge $Q^{(k)}$)
for the same number $n_u$ of local updating steps.
The plateaus showed after a few heating steps 
by data plotted as function of $n_u$ give the desired estimates
of $Z^{(k)}$.
Similarly, the background signal $M^{(k)}$ can be estimated 
by measuring $\chi^{(k)}$ on ensembles of configurations
constructed by heating the flat configuration. Again, 
the plateaus of data as function of $n_u$, if observed,
should provide estimates of $M^{(k)}$~\cite{heating,ACDGV}.

We performed our simulations on a $100^2$ lattice.
This lattice size should be sufficiently large in order
to estimate $Z$ and $M$ even at large $\beta$,
in that they are expected to be short ranged
(scaling with $a$), and therefore have 
very small finite size effects.
On the other hand, in order to determine $\chi_t$, one
must take lattices  with  $L\gg \xi$.
In the following we describe the main results we have obtained
for $N=4$ and $N=10$ using the lattice
action $S^{(g)}_L$ (cfr. Eq.~(\ref{basic})).

For $\beta\simeq 1$
(where the correlation length is sufficiently large to expect scaling
to hold~\cite{CPN}), 
the multiplicative renormalization $Z^{(k)}$ turns out to be very close
to one for small values of $k$ already.
For example for $N=4$ and at $\beta=1.25$ (which corresponds to 
$\xi_G\simeq 28$) we found  $Z^{(1)}\simeq 0.96$,
$Z^{(2)}\simeq 0.98$, $Z^{(3)}\simeq 0.99$, ...,
suggesting a smooth limit to one for $k\rightarrow\infty$.
For comparison we mention that the polynomial topological charge
density definition considered in Ref.~\cite{CPN},
\begin{equation}
q^{(p)}_{n} = -{i\over 2\pi}\sum_{\mu\nu} \epsilon_{\mu\nu}
   {\rm Tr}\left[ P_n\Delta_\mu P_n 
   \Delta_\nu P_n \right]
\label{localq1}
\end{equation}
(where $\Delta_\mu P(x) = \case1/2 [P_{n{+}\mu} - P_{n{-}\mu}]$),
has $Z^{(p)}\simeq 0.42$ at this value of $\beta$.
At larger $N$ $Z^{(k)}$ get closer to one, as expected from their
dependence on $N$. For example, at $N=10$ and $\beta=1.0$ 
(corresponding to $\xi_G\simeq 17$) we found 
$Z^{(1)}\simeq 0.985$,
$Z^{(2)}\simeq 1.000$, $Z^{(3)}\simeq 1.0000$, ...,
and $Z^{(p)}\simeq 0.35$.

When using the action (\ref{basic}), at $N=4$ and $\beta\simeq 1.25$ 
the geometrical definition (\ref{qgeomP}) is affected
by spurious contributions from short-ranged lattice structures,
which turn out not to be negligible in order to determine $\chi_t$
from $\chi_g=\case{1}{V}\langle Q_g^2\rangle$~\cite{CPN}.
This fact emerges clearly also by observing the behavior
of $\chi_g$ in the heating of the flat configuration,
in that the initial global geometrical charge (which is zero)
appears to be soon modified by the local thermalization
process even if the correlation length is rather large, as it occurs at
$\beta\simeq 1.25$.
In the same heating process 
data of $\chi^{(k)}$ appear strongly correlated to those of
$\chi_g$, indicating that the above-mentioned short-ranged
configurations contribute somehow to $\chi^{(k)}$ too (even for small $k$).
In the analysis of $\chi^{(k)}$ such spurious contributions
may be interpreted as renormalization effects,
but at these values of $N$ and $\beta$
 we were not able to estimate them
by using the heating method.
It is possible that in this case
the required sharp distinction of the
lattice modes contributing to the different terms
in Eq.~(\ref{mixingsusccpn}) does not occur, or
it is not sufficient to make the heating method work.

In a sense,
the so-called lattice defect contributions to the geometrical
definition $\chi_g$ may be seen as the limit $k\rightarrow\infty$
of the mixings in $\chi^{(k)}$,
which does not seem to vanish at $N=4$ and values of
$\beta$ corresponding to $\xi_G\lesssim 10^2$.
Their effect probably disappears in the large-$\beta$ limit.
Indeed
at $\beta=1.6$, where the correlation length should
be about one order of magnitude larger than at $\beta=1.25$, 
we found no trace of
short-ranged topological defects in the heating procedure.
We got the
following  estimates of the mixing contribution $M^{(k)}$ 
to $\chi^{(k)}$:
$M^{(1)}\simeq 3\times 10^{-7}$,
$M^{(2)}\lesssim 5\times 10^{-8}$,
$M^{(3)}\lesssim 10^{-8}$,...,
while for the lattice susceptibility associated with 
the operator (\ref{localq1}) 
we found $M^{(p)}\simeq 7 \times 10^{-6}$.
$\beta\gtrsim 1.6$, and therefore  $\xi\gtrsim  10^2$, may then
be sufficiently large for the geometrical definition
(\ref{qgeomP}) to be effective  in the determination of $\chi_t$.
We also quote some data at $\beta=2$ where the correlation
length should be about two order of magnitude
larger than at $\beta\simeq 1.25$,
we found $M^{(1)}\simeq 2.5\times 10^{-8}$,
$M^{(2)}\lesssim 2\times 10^{-10}$,
$M^{(3)}\lesssim 10^{-11}$,...,
and $M^{(p)}\simeq 2.9 \times 10^{-6}$.
Notice that in order to determine $\chi_t$ at $\beta=1.6$ one
 would need to perform simulations on a very large lattice
with $L\gg \xi$, and therefore $L>10^3$ in order
to avoid sizeable finite size effects.
Of course the onset of scaling for $\chi_t$
depends on the lattice action. In this respect 
actions better than (\ref{basic}) may exist. 
A substantial improvement with respect
to action (\ref{basic}) has been observed when using its Symankik
tree-improvement~\cite{CPN}. 
These results are consistent
with the picture outlined in the previous subsection.

As suggested by Eq.~(\ref{pp}), things get improved with increasing
$N$. For $N\gtrsim 10$ (using the action (\ref{basic}))
the geometrical definition (\ref{qgeomP}) seems to provide a good 
estimator of $\chi_t$ already at reasonable values of the correlation
length~\cite{CPN}. For example at $N=10$,
$\chi_g\xi_G^2$ shows good scaling for $\xi_G\gtrsim
10$~\cite{CPN}, and for shorter and shorter $\xi_G$
at larger values of $N$.
At $N=10$ and $\beta=1.0$, the heating procedure 
provided the following estimates of the background signal $M^{(k)}$ 
in $\chi^{(k)}$:
$M^{(1)}\simeq 0.5\times 10^{-6}$,
$M^{(2)}\lesssim 0.5\times 10^{-7}$,
$M^{(3)}\lesssim 0.5\times 10^{-8}$,
$M^{(4)}\lesssim 10^{-9}$ ...,
and $M^{(p)}\simeq 0.7\times 10^{-5}$ for the operator (\ref{localq1}). 
The mixings
in Eq.~(\ref{mixingsusccpn}) rapidly disappear with increasing $k$.
The sequence of $\chi^{(k)}$ approaches $\chi_g$ which in this case
appears to be free of lattice artifact contributions.    
Taking into account that for $N=10$ we have
$\chi_t\xi_G^2\simeq 0.017$~\cite{CPN},
at $\beta=1.0$ renormalization effects turn out to be very small and
negligible for $k=1$ already (in the evaluation of $a^2\chi_t$
renormalizations for $k=1$ lead to corrections of about one per cent).

From the above discussion it seems that,
at least when using the action (\ref{basic}), little practical
improvement is achieved by the use of the 
operators defined by the sequences $q^{(k)}_n$.
Indeed at low $N$ (and at least for $\beta\simeq 1$)
they appear to be sensitive to short-ranged lattice defects,
whose contributions to renormalization effects
seem as difficult to evaluate as 
for the corresponding geometrical definition
(the heating method apparently fails in these cases).
The use of standard polynomial or smeared field theoretical operators 
(smeared operators similar to those defined
in Ref.~\cite{CDHV} and with the same
features can be easily constructed for 
$CP^{N-1}$ models) and 
cooling method~\cite{coolingmethod} appears more convincing
when this phenomenon occurs.
When the geometrical definition provides a good estimator of topological
activity, i.e. at large $N$, the behavior of $q^{(k)}$ is that 
formally predicted by perturbation theory.
But in this case it would be more
convenient to use the corresponding
geometrical definition, whose use is further justified by our
analysis.
However when the renormalization effects do not tend
to vanish (at fixed $\beta$) along the sequence, we cannot take for
granted that the physical predictions based on the use of the
corresponding geometrical definition will be correct.

In order to construct improved field theoretical
operators one may put together both the idea of smearing the 
operators~\cite{CDHV} and that of defining sequences approaching
geometrical definitions. This may be achieved
by constructing sequences of operators
in terms of  smeared fields, instead of lagrangian fields.
This should provide optimal field theoretical
operators, i.e. with a more effective suppression
of the renormalization effects, and corresponding
geometrical limits probably less sensitive to 
unphysical short-distance lattice topological defects.
We mention that for 4-d $SU(2)$ gauge theory a geometrical charge
defined on appropriate blocked variables has been tested~\cite{Teper2}.
Unlike the original geometrical charges which turned out to be affected
by dislocations, the blocked definition produced
a topological susceptibility consistent with that obtained by
alternative cooling and field theoretical methods.

\section{Construction of the sequence for $SU(N)$ Yang Mills theories}
\label{sec3}
 
We sketch an extension of the above study
to $SU(N)$ Yang Mills
theories in four dimensions. Among the different geometrical definitions
of lattice topological charge available in the literature,
we consider that one proposed by L\"uscher in Ref.~\cite{Luscher}.
In the following we will heavily refer to this work.
This geometrical definition can be written as a sum of local terms
$Q_g=\sum_n q_n$, where $q_n$ is a gauge invariant function of the
link variables.
We show that it is possible to define a sequence of operators $q_n^{(k)}$ 
sharing the following properties.

(i) For each $k$, $q_n^{(k)}$ is a gauge invariant, polynomial function
of the link variables of the single elementary hypercube $c(n)$, possessing
the appropriate (classical) continuum limit;

(ii) The operators $q_n^{(k)}$ tend, at least formally, to $q_n$
for $k\rightarrow\infty$.

$q_n$ is obtained by the lattice fields of the hypercube
$c(n)$ through a complicated interpolation procedure.
An essential ingredient is the raising of link 
variables to fractional powers
(this is the non-analitycal step in the definition), according
to the following prescription. 
For  $u \in SU(N)$, setting
\begin{equation}
u=\exp (i g\omega_a \lambda_a),
\end{equation}
one may define the fractional power $u^y$, $0 \leq y \leq 1$, as
\begin{equation}
 u^y = \exp (i y g\omega_a \lambda_a)
\end{equation}
(for convenience, we explicitly introduced the bare 
coupling costant $g$, in terms of which the perturbative weak 
coupling expansion is defined).

We construct a sequence of gauge invariant analytical
operators $q^{(k)}_n$ which differ from the geometrical definition $q_n$
by $O(g^{k+1})$, i.e.
\begin{equation}
q^{(k)}_n=q_n+O\left(g^{k+1}\right). 
\label{qkqk}
\end{equation}
To this purpose, we consider a polynomial 
approximation $u_{[j]}(y)$ of $u^y$, of degree $j$ in $u$ and $u^\dagger$,
so that  
\begin{equation}
u^y = u_{[j]}(y) + O(g^{2j+1}). 
\end{equation}
The polynomial functions $u_{[j]}(y)$ can be easily
constructed. 
We then make 
the appropriate substitutions $u^y \rightarrow u_{[j]}(y)$
in the expression of L\"uscher's topological charge density
$q_n$, so to obtain $q_n^{(k)}$ as an approximation 
$O(g^{k+1})$ of $q_n$
that contains polynomials of minimum degree in the link variables.

The lattice susceptibilities  
$\chi^{(k)}=\sum_n \langle q_n^{(k)}q_0^{(k)}\rangle$ 
should be related to the continuum topological susceptibility
$\chi_t$ by
\begin{equation}
\chi^{(k)}(\beta)  =   
a^4 Z^{(k)}(\beta)^2\chi_t + M^{(k)}(\beta),
\label{mixingsuscQCD} 
\end{equation}
where
\begin{equation}
M^{(k)}(\beta)=P^{(k)}(\beta)\left<I\right>  + a^4  
A^{(k)}(\beta) \left<T \right> \ + \ O(a^6).  \label{mixingsuscQCD2} 
\end{equation}
Assuming non-renormalization for the geometrical definition $q_n$,
and using Eq.~(\ref{qkqk}) one can infer that
($\beta = \frac{2 N}{g^2}$):
\begin{eqnarray}
Z^{(k)}(\beta) &=& 1 + O\left( {1\over \beta^l} \right), \\
P^{(k)}(\beta) &=& O\left( {1\over \beta^{l+2}}\right),
\end{eqnarray}
where
\begin{eqnarray}
l&=&{k\over 2} \qquad\qquad {\rm for \;\;\; even \;\;\;}k,\nonumber \\
l&=&{k+1\over 2} \qquad\;{\rm for \;\;\; odd \;\;\;}k.
\end{eqnarray}
An analysis of the behavior
of the coefficients $z^{(k)}$ and $p^{(k)}$
of the leading non-trivial
order in $Z^{(k)}$ and $P^{(k)}$ 
can be more easily carried out for 
even $k$, for which all the corresponding diagrams are
substantially tadpoles.
A rough estimate of the behavior of $p^{(k)}$ and $z^{(k)}$
may be obtained by 
counting all the contractions to form $k/2$ tadpoles:
\begin{equation}
p^{(k)} \sim z^{(k)} \sim {k!\over (k/2)!}.
\label{est}
\end{equation} 
Here the dependence on $N$ has been overlooked, in that
all contractions have been considered as giving the same
contribution. This is true only in the case of
 commuting generators, as in the case of $U(1)$ gauge theory.
Neverthless Eq.~(\ref{est}) should give an idea of the behavior
at large $k$, at least for not too large $N$. 
On the other hand, at large-$N$  contractions of 
non-sequential generators in the traces are suppressed
by powers of $1/N$, so that the 
contractions contributing 
 at $N=\infty$ are
considerably reduced, leading probably to
$p^{(k)} \sim z^{(k)}\sim O(1)$.

Then, as already observed for the sequences constructed within the
$CP^{N-1}$ models,
the lowest-order renormalizations are proportional to 
higher and higher powers of $1/\beta$. However, at finite $N$
the corresponding coefficients of the leading non-trivial order
grow so fast with $k$ that the convergence to 
zero of the renormalization functions
cannot be uniform for $\beta \rightarrow \infty$. 
Again, with increasing $N$  renormalization effects should
be further suppressed in the sequence, suggesting that
at least at large-$N$ 
the geometrical charge should be free from dislocations.

\acknowledgments

It is a pleasure to thank B. All\'es, A.~Di~Giacomo,
 and H. Panagopoulos
for useful and stimulating discussions.



\end{document}